\documentclass[letterpaper]{article}

\usepackage{aaai}

\usepackage{times}
\usepackage{helvet}
\usepackage{courier}
\usepackage{amsmath}
\usepackage{amssymb}
\usepackage{graphicx}

\DeclareMathOperator*{\BR}{BR}
\DeclareMathOperator*{\CBR}{CBR}
\DeclareMathOperator*{\depth}{depth}

\def\etal{\textit{et al.}}
\newcommand{\ie}{{\it i.e., }}

\DeclareMathOperator*{\argmax}{\arg\!\max}
\newtheorem{theorem}{Theorem}
\newtheorem{lemma}{Lemma}
\newtheorem{defn}{Definition}
\newenvironment{proof}{\paragraph{Proof}}{\strut~\hfill $\Box$\medskip}

\newif\ifarxiv
\arxivtrue

\begin{document}

\nocopyright
\title{Solving Imperfect Information Games Using Decomposition}
\author{Neil Burch, Michael Johanson and Michael Bowling \\
Computing Science Department, University of Alberta \\
\{nburch,johanson,mbowling\}@ualberta.ca}

\maketitle

\begin{abstract}
Decomposition, \ie independently analyzing possible subgames, has
proven to be an essential principle for effective decision-making in
perfect information games.  However, in imperfect information games,
decomposition has proven to be problematic.  To date, all proposed
techniques for decomposition in imperfect information games have
abandoned theoretical guarantees.  This work presents the first
technique for decomposing an imperfect information game into subgames
that can be solved independently, while retaining optimality
guarantees on the full-game solution.  We can use this technique to
construct theoretically justified algorithms that make better use of
information available at run-time, overcome memory or disk limitations
at run-time, or make a time/space trade-off to overcome memory or disk
limitations while solving a game.  In particular, we present an
algorithm for subgame solving which guarantees performance in the
whole game, in contrast to existing methods which may have unbounded
error.  In addition, we present an offline game solving algorithm,
CFR-D, which can produce a Nash equilibrium for a game that is larger
than available storage.
\end{abstract}

\section{Introduction}
\label{sec:introduction}
A game solving algorithm takes the description of a game and computes or approximates an optimal strategy (\ie a Nash
equilibrium) for playing the game.  Perfect information games, such as
checkers, where game states are entirely public, have historically
been more tractable to solve than imperfect information games, such as
poker, where some information about the game state is hidden from one
or more players.  The main reason is that perfect information games
can easily be partitioned into subgames that can be solved
independently, producing strategy fragments that can be combined to
form an optimal strategy for the entire game.  

Reasoning about subgames independently has two highly desirable
properties.  First, decomposition can allow large savings
in the memory required to solve a game.  If we split a game with $S$
states into subgames half-way to the end of the game, we end up with
$\mathcal{O}(S)$ subgames each of size $\mathcal{O}(\sqrt{S})$: a single
``trunk'' spanning from the start of the game to the split depth, plus
a number of subgames.  If we only need to reason about a 
single subgame at a time, then we use an amount of storage on the
order of $\mathcal{O}(\sqrt{S})$ instead of $\mathcal{O}(S)$.  The subgame pieces can also be recursively
decomposed, so that in perfect information games that are no more than
$D$ actions long, a game solving algorithm like depth-first
iterative-deepening~\cite{korf85} uses only $\mathcal{O}(D)$ memory.
 Second, we do not need to
store the complete strategy, which may be to large to store, but rather can recompute the subgame
strategies as needed.  
As a result such perfect information decomposition algorithms are effectively not limited by space, and with
sufficient time can solve extremely large games.
For example, checkers with $5 \times 10^{20}$ states has
been solved~\cite{07science-checkers} both in terms of the game's
value and an optimal Nash equilibrium strategy.

In imperfect information games, there are currently no methods for
solving, or re-solving, subgames with a guarantee that
the subgame strategies can be combined into an equilibrium for
the whole game.  State-of-the-art algorithms are all limited to
comparatively small problems where the complete strategy fits in
available space.
As a result, 2-Player Limit Texas Hold'em Poker, with $9 \times
10^{17}$ game states, is smaller than checkers but has not been
solved despite considerable recent interest in the game.  Computing an optimal
strategy for this game would require hundreds of terabytes of memory using a
state-of-the-art game solving algorithm.

In this paper we present, for the first time, two methods which safely use
decomposition in imperfect information games.  We give a new definition of
subgames which is useful for imperfect information games, and a method
for re-solving these subgames which is guaranteed to not increase the
exploitability (\ie suboptimality) of a strategy for the whole game.  
We also give a general method called CFR-D for computing an
error-bounded approximation of a Nash equilibrium through decomposing
and independently analyzing subgames of an imperfect information game.
Finally, we give experimental results comparing our new methods to
existing techniques, showing that the prior lack of theoretical bounds
can lead to significant error in practice.

\section{Notation and Background}
\label{sec:background}
An extensive-form game is a model of sequential interaction of one or
more agents or players.  Let $P$ be the set of players.  
Let $H$ be the set of all
possible game states, represented as the history of actions taken from
the initial game state $\varnothing$.  The state $h \cdot a \in H$ is a child 
of the state $h$, $h$ is the parent of $h \cdot a$, and $h'$ is a 
descendant of $h$ or $h \sqsubset h'$ if $h$ is
any strict prefix of $h'$. Let $Z$ be the set of all terminal states.  For
each non-terminal state $h$, $A(h)$ gives the set
of legal actions, and $P(h) \in P \cup \{c\}$ gives the player to act,
where $c$ denotes the ``chance player'', which represents stochastic
events outside of the players' control.  $\sigma_c(h, a)$ is the
probability that chance will take action $a \in A(h)$ from state
$h$, and is common knowledge.  $H_p$ is the set of all states $h$ such that
$P(h)=p$.  For every $z \in Z$, $u_p(z) \in \Re$ gives the payoff
for player $p$ if the game ends in state $z$.  If $p=\{1,2\}$ and
$u_1(z) + u_2(z) = 0$ for all $z\in Z$, we say the game is two-player,
zero-sum.

The information structure of the game is described by information
sets for each player $p$, which form a partition $\mathcal{I}_p$ of $H_p$.  For any
information set $I \in \mathcal{I}_p$, any two states $h,j \in I$ are
indistinguishable to player $p$.  Let $I(h)$ be the information set in $\mathcal{I}_p$ which
contains $h$.
A behaviour strategy $\sigma_p \in
\boldsymbol{\Sigma}_p$ is a function $\sigma_p(I,a) \in \Re$ which
defines a probability distribution over valid actions for every
information set $I \in \mathcal{I}_p$.  We will say
$\sigma_p(h,a)=\sigma_p(I(h),a)$, since a player cannot act differently depending on
information they did not observe.  
Let $Z(I) = \{ z \in Z \mbox{~s.t.~} z \sqsupset h
\in I\}$ be the set of all terminal states $z$ reachable from some
state in information set $I$.  We can also consider the terminal states
reachable from $I$ after some action $a$, stated as $Z(I,a) = \{ z \in Z \mbox{~s.t.~}  z \sqsupset h \cdot a, h \in I\}$.

In games with perfect recall, any two states $h$ and $j$ in an
information set $I \in \mathcal{I}_p$ have the same sequence of
player $p$ information sets and actions.  Informally, perfect recall means
that a player does not forget their own actions or any information observed
 before making those actions.  As a result, for any
$z \in Z(I)$ there is a unique state $h \in I$ such that $h
\sqsubset z$, which we write $z[I]$.  This paper focuses exclusively
on two player, zero-sum, perfect recall games.

A strategy profile $\sigma \in \boldsymbol{\Sigma}$ is a tuple of
strategies, one for each player.  Given $\sigma$, it is useful to
refer to certain products of probabilities.  Let $\pi^\sigma(h) =
\prod_{j\cdot a \sqsubseteq h}{\sigma_{P(j)}(j,a)}$, which gives the joint probability of
reaching $h$ if all players follow $\sigma$.  We use $\pi^\sigma_p(h)$ to
refer to the product of only the terms where $P(h) = p$, and
$\pi^\sigma_{-p}(h)$ to refer to the product of terms where $P(h) \ne
p$.
Note that in games with perfect recall, for all states $h$, $h'$ in $I \in \mathcal{I}_p$, $\pi_p(h)=\pi_p(h')$, so we can also speak of $\pi_p(I)$.
We use $\pi^\sigma(j,h)$ to refer to the product of terms from $j$
to $h$, rather than from $\varnothing$ to $h$.  If we replace the
whole strategy for player $p$ by a new strategy $\sigma'_p$, we will
call the resulting profile $\langle\sigma_{-p}, \sigma'_p\rangle$.  
Finally,
$\sigma_{[S \gets \sigma']}$ is the strategy that is equal to $\sigma$
everywhere except at information sets in $S$, where it is equal to
$\sigma'$.

Given a strategy profile $\sigma$, the expected utility $u_p^{\sigma}$
to player $p$ if all players follow $\sigma$ is
$\sum_Z{\pi^{\sigma}(z)u_p(z)}$.  The expected utility $u_p^{\sigma}(I,a)$ of
taking an action at an information set is $\sum_{z \in
Z(I,a)}{\pi^\sigma(z)u_p(z)}$.  In this paper, we will frequently use
a variant of this expected value called counterfactual value:
$v^\sigma_p(I,a) = \sum_{z \in
Z(I,a)}{\pi^\sigma_{-p}(z)\pi^\sigma_p(z[I] \cdot a,z)u_p(z)}$.
Informally, the counterfactual value of $I$ for player $p$ is the
expected value of reaching $I$ if $p$ plays to reach $I$.

A best response $\BR_p(\sigma) =
\argmax_{\sigma'_p \in \boldsymbol{\Sigma}_p}
u_p^{\langle\sigma_{-p}, \sigma'_{p}\rangle}$ is a strategy for $p$
which maximises $p$'s value if all other player strategies remain
fixed.  A Nash equilibrium is a strategy profile where all strategies
are simultaneously best responses to each other, 
and an $\epsilon$-Nash equilibrium 
is a profile where the expected value for each player is within
$\epsilon$ of the value of a best response strategy.  In two-player, zero-sum games, the expected utility of any Nash
equilibrium is a game-specific constant, called the game
value.  In a two-player zero-sum game,
we use the term exploitability to refer to a profile's average
loss to a best response across its component strategies.  A Nash
equilibrium has an exploitability of zero. 

A counterfactual best response $\CBR_p(\sigma)$ is a strategy where
$\sigma_p(I, a) > 0$ if and only if $v_p(I, a) \ge \max_b v_p(I, b)$,
so it maximizes counterfactual value at every information set.
$\CBR_p$ is necessarily a best response, but
$\BR_p$ may not be a counterfactual best response as it may choose
non-maximizing actions where $\pi_p(I) = 0$.  The well known recursive bottom-up technique of
constructing a best response generates a counterfactual best response.

\section{Decomposition into Subgames}
In this paper we introduce a new refinement on the concept of a
subgame.  A subgame, in a perfect information
game, is a tree rooted at some arbitrary state: a set of states closed
under the descendant relation.  The state-rooted subgame definition is
not as useful in an imperfect information game because the tree cuts
across information set boundaries: for any state $s$ in the tree,
there is generally at least one state $t \in I(s)$ which is not in the
tree.

To state our refined notion of subgame it is convenient to extend the concept of an information set.
$I(h)$ is defined in terms of the states which player $p=P(h)$ cannot
distinguish.  We would also like to partition states where player $p$
acts into those which player $p' \ne p$ cannot distinguish.  We use
the ancestor information sets to construct $I_{p'}(h)$, the augmented
information set for player $p'$ containing $h$.  Let $H_{p'}(h)$ be
the sequence of player $p'$ information sets reached by player $p'$ on
the path to $h$, and the actions taken by player $p'$.  Then for two
states $h$ and $j$, $I_{p'}(h) = I_{p'}(j) \iff H_{p'}(h) =
H_{p'}(j)$.

We can now state the following definition of a subgame:
\begin{defn}An imperfect information subgame is a forest of trees, closed under
both the descendant relation and membership within augmented information sets
for any player.\end{defn}

The imperfect information subgame is a forest rooted at a set of
augmented information sets.  If state $s$ is in the subgame, and $s \sqsubseteq
t$ or $s,t \in I_p$ for any information set $I_p$, then state $t$ is
also in the subgame.  Note that the root of the subgame will not
generally be a single information set, because different players will
group states into different information sets.  We use augmented
information sets in this definition because we wish to preserve the
information partitioning at the root of the subgame.  For example, say
player one can distinguish states $s$ and $t$ where player one is
acting, and player two can distinguish their descendants, but not
their ancestors.  If we did not use augmented information sets, a
subgame could include $s$ and not include $t$.  We use augmented
information sets to rule out this case.  In a perfect information
game, our definition is equivalent to the usual definition of a
subgame, as information sets all contain a single state.

\begin{figure}
  \centering\includegraphics[width=0.8\hsize]{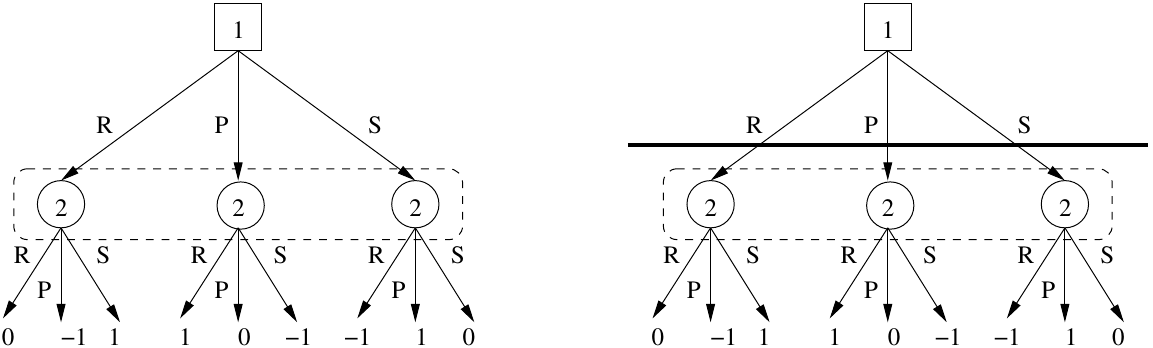}
  \caption{Left: rock-paper-scissors.  Right: rock-paper-scissors split into trunk and one subgame.}
  \label{fig:rps}
\end{figure}

We will use the game of rock-paper-scissors as a running example in
this paper.  In
rock-paper-scissors, two players simultaneously choose rock, paper, or
scissors.  They then reveal their choice, with rock beating scissors,
scissors beating paper, and paper beating rock.  The simultaneous
moves in rock-paper-scissors can be modeled using an extensive form
game where one player goes first, without revealing their action, then
the second player acts.  The extensive form game is shown on the left
side of Figure~\ref{fig:rps}.  The dashed box indicates the
information set $I_2=\{R,P,S\}$ which tells us player two does not
know player one's action.

On the right side of Figure~\ref{fig:rps}, we have decomposed the game
into two parts: a trunk containing state $\varnothing$ and a single
subgame containing three states $R$, $P$, and $S$.  In the subgame,
there is one player two information set $I_2=\{R,P,S\}$ and three
augmented player one information sets $I_1^R=\{R\}$, $I_1^P=\{P\}$,
and $I_1^S=\{S\}$.

\section{Subgame Strategy Re-Solving}
\label{sec:subgame}
In this section we present a method of re-solving a subgame, using
some compact summary information retained from a previous strategy in
this subgame.  The novel property of this method is a bound on the exploitability of
the combined trunk and new subgame strategy in the whole game.  This
sort of
re-solving problem might be useful in a number of situations.  For example, we
might wish to move a strategy from some large machine to one with very
limited memory.  If we can re-solve subgame strategies as needed, then
we can discard the original subgame strategies to save space.  Another
application occurs if the existing strategy is suboptimal, and we wish
to find a better subgame strategy with a guarantee that the new
combined trunk and subgame strategy does at least as well as the
existing strategy in the worst case.  An additional application of
space reduction while solving a game is presented later in this paper.

First, we note that it is not sufficient to simply re-solve the
subgame with the assumption that the trunk policy is fixed.  When
combined with the trunk strategy, multiple subgame solutions may
achieve the same expected value if the opponent can only
change their strategy in the subgame, but only a subset of these
subgame strategies will fare so well against a best
response where the opponent can also change their strategy in the trunk.

Consider the rock-paper-scissors example.  Let's say we started with
an equilibrium, and then discarded the strategy in the subgame.  In
the trunk, player one picks uniformly between $R$, $P$, and $S$.  In
the subgame, player one has only one possible (vacuous) policy: they
take no actions.  To find an equilibrium in the subgame, player two
must pick a strategy which is a best response to the empty player one
policy, given the probability of $\frac{1}{3}$ for $R$, $P$, and $S$
induced by the trunk strategy.  All actions have an expected utility
of $0$, so player two can pick an arbitrary policy.  For example,
player two might choose to always play rock.  Always playing rock
achieves the game value of $0$ against the combined trunk and subgame
strategy for player one, but gets a value of $-1$ if player one
switched to playing paper in the trunk.

Our new method of re-solving subgames relies on summarising a subgame
strategy with the opponent's counterfactual values $v_{opp}(I)$ for
all information sets $I$ at the root of the subgame.  $v_{opp}(I)$
gives the ``what-if'' value of our opponent reaching the
subgame through information set $I$, if they changed their strategy so
that $\pi_{opp}(I)=1$.  In rock-paper-scissors, the player one
counterfactual values for $R$, $P$, and $S$ are all $0$ in the
equilibrium profile.  When player two always played rock in the
example above, the player one counterfactual values for $R$, $P$, and
$S$ were $0$, $1$, and $-1$ respectively.  Because the counterfactual
value for $P$ was higher than the original equilibrium value of $0$,
player one had an incentive to switch to playing $P$.  That is, they
could change their trunk policy to convert a larger ``what-if''
counterfactual value into a higher expected utility by playing $P$.

If we generate a subgame strategy where the opponent's best response
counterfactual values are no higher than the opponent's best response
counterfactual values for the original strategy, then the
exploitability of the combined trunk and subgame strategy is no higher
than the original strategy.  From here on, we will assume, without
loss of generality, that we are re-solving a strategy for player $1$.

\begin{theorem}
\label{theorem:resolve-subgames}
Given a strategy $\sigma_1$, a subgame $S$, and a
re-solved subgame strategy $\sigma^S_1$, let $\sigma'_1
= \sigma_{1,[S \gets \sigma^S_1]}$ be the combination of $\sigma_1$
and $\sigma^S_1$. If
$v_2^{\langle\sigma'_1,\CBR(\sigma'_1)\rangle}(I) \le
v_2^{\langle\sigma_1,\CBR(\sigma_1)\rangle}(I)$ for all information
sets $I$ at the root of subgame $S$, then
$u_2^{\langle\sigma'_1,\CBR(\sigma'_1)\rangle}
\le u_2^{\langle\sigma_1,\CBR(\sigma_1)\rangle}$.
\end{theorem}

\ifarxiv
A proof of Theorem~\ref{theorem:resolve-subgames} is given in the appendix.
\else
A proof of Theorem~\ref{theorem:resolve-subgames} is given in the appendix of ~\cite{13arxiv-cfrd}.
\fi

\begin{figure}
  \centering\includegraphics[width=0.8\hsize]{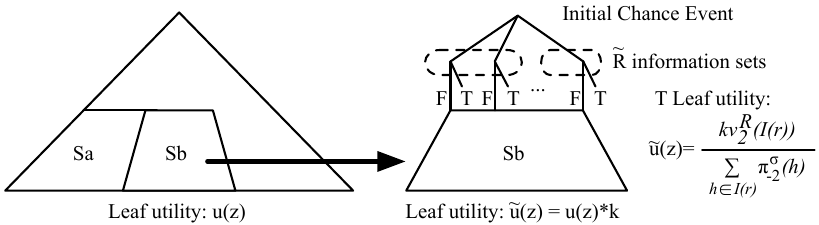}
  \caption{Construction of the Re-Solving Game}
  \label{fig:recovery-game}
\end{figure}

To re-solve for a strategy in a subgame, we will construct the
modified subgame shown in Figure~\ref{fig:recovery-game}.  We will
distinguish the re-solving game from the original game by using a
tilde ($\;\tilde{}\;$) to distinguish states, utilities, or strategies
for the re-solving game.  The basic construction is that each state
$r$ at the root of the original subgame turns into three states:
a $p_2$ choice node $\tilde{r}$, a terminal state $\tilde{r} \cdot T$,
and a state $\tilde{r} \cdot F$ which is identical to $r$.  All other
states in the original subgame are directly copied into the re-solving
game.  We must also be given $v^R_2(I) \equiv
v^{\langle\sigma_1,\CBR(\sigma_1)\rangle}_2(I)$ and
$\pi^\sigma_{-2}(I)$ for all $p_2$ information sets
$I \in \mathcal{I}_2^R$ at the root of the subgame.

The re-solving game beings with an initial chance node which leads to
states $\tilde{r} \in
\tilde{R}$, corresponding to the probability of reaching state
$r \in R$ in the original game.  Each state $\tilde{r} \in \tilde{R}$
occurs with probability $\pi^\sigma_{-2}(r) / k$, where the constant
$k = \sum_{r \in R} \pi^\sigma_{-2}(r)$ is used to ensure that the
probabilities sum to $1$.  $\tilde{R}$ is partitioned into information
sets $\mathcal{I}_2^{\tilde{R}}$ that are identical to the information
sets $\mathcal{I}_2^{R}$.

At each $\tilde{r} \in \tilde{R}$, $p_2$ has a binary choice of $F$ or
$T$.  After $T$, the game ends.  After $F$, the game is the same as
the original subgame.  All leaf utilities are multiplied by $k$ to
undo the effects of normalising the initial chance event.  So, if
$\tilde{z}$ corresponds to a leaf $z$ in the original subgame,
$\tilde{u}_2(\tilde{z}) = ku_2(z)$.  If $\tilde{z}$ is a terminal
state after a $T$ action, $\tilde{u}_2(\tilde{z}) =
\tilde{u}_2(\tilde{r} \cdot T) = kv^R_2(I(r)) / \sum_{h \in
  I(r)}\pi^\sigma_{-2}(h)$.  This means that for any $I \in
\mathcal{I}_2^{\tilde{R}}$, $\tilde{u}_2(I \cdot T) = v^R_2(I)$, the original counterfactual best response value of $I$.

No further construction is needed.  If we solve the proposed game to
get a new strategy profile $\tilde{\sigma^*}$, we can directly use
$\tilde{\sigma^*}_1$ in the original subgame of the full game.  To see
that $\sigma^*_1$ achieves the goal of not increasing the
counterfactual values for $p_2$, consider $\tilde{u}_2(I)$ for
$I \in \mathcal{I}_2^{\tilde{R}}$ in an equilibrium profile for the
re-solving game.  $p_2$ can always pick $T$ at the initial choice to
get the original counterfactual values, so $\tilde{u}_2(I) \ge
v^R_2(I)$.  Because $v_2^R$ comes from $\langle\sigma_1,\CBR(\sigma_1)\rangle$,
$\tilde{u}_2(I) \le v^R_2(I)$ in an equilibrium. So, in a solution
$\tilde{\sigma^*}$ to the re-solving game, $\tilde{u}_2(I) = v^R_2(I)$,
and $\tilde{u}^{\langle\tilde{\sigma^*}_1,\CBR(\tilde{\sigma^*}_1)\rangle}_2(I \cdot
F) \le v^R_2(I)$.  By construction of the re-solving game, this
implies that $v^{\langle\sigma^*_1,\CBR(\sigma^*_1)\rangle}_2(I) \le v^R_2(I)$.

If we re-solve the strategy for both players at a subgame, the
exploitability of the combined strategy is increased by no more than
$(|\mathcal{I}^{R_S}|-1)\epsilon_{S} + \epsilon_{R}$, where
$\epsilon_{R}$ is the exploitability of the subgame strategy in the
re-solving subgame, $\epsilon_{S}$ is the exploitability
of the original subgame strategy in the full game, and
$|\mathcal{I}^{R_S}|$ is the number of information sets for both
players at the root of a subgame.
\ifarxiv
This is proved in Theorem~\ref{theorem:resolved-cfrd} of the appendix.
\else
This is proved as Theorem 3 in~\cite{13arxiv-cfrd}.
\fi

\section{Generating a Trunk Strategy using CFR-D}
\label{sec:trunk}

CFR-D is part of the family of counterfactual regret minimisation
(CFR) algorithms, which are all efficient methods for finding an
approximation of a Nash equilibrium in very large games.  CFR is an
iterated self play algorithm, where the average policy across all
iterations approaches a Nash equilibrium~\cite{07nips-cfr}.  It has
independent regret minimisation problems being simultaneously updated
at every information set, at each iteration.  Each minimisation
problem at an information set $I \in \mathcal{I}_p$ uses immediate
counterfactual regret, which is just external regret over
counterfactual values: $R^T(I)=\displaystyle \max_{a \in A(I)}
\sum_t v^{\sigma^t}_{P(I)}(I,a) - \sum_{a'}
\sigma^t(I,a')v^{\sigma^t}_{P(I)}(I,a')$. 
The immediate counterfactual regrets place an upper bound on the
regret across all strategies, and an $\epsilon$-regret strategy
profile is a $2\epsilon$-Nash equilibrium~\cite{07nips-cfr}.

Using separate regret minimisation problems at each information set
makes CFR a very flexible framework.  First, any single regret
minimisation problem at an information set $I$ only uses the
counterfactual values of the actions.  The action probabilities of the
strategy profile outside $I$ are otherwise irrelevant.  Second, while
the strategy profile outside $I$ is generated by the other
minimisation problems in CFR, the source does not matter.  Any
sequence of strategy profiles will do, as long as they have low
regret.

The CFR-BR algorithm~\cite{12aaai-cfrbr} uses these properties, and
provided the inspiration for the CFR-D algorithm.  The game is split
into a trunk and a number of subgames.  At each iteration, CFR-BR
uses the standard counterfactual regret minimisation update for both
players in the trunk, and for one player in the
subgames.  For the other player, CFR-BR constructs and uses
a best response to the current CFR player strategy in each subgame.

In our proposed algorithm, CFR-D, we use a counterfactual best
response in each subgame for both players.  That is, at each
iteration, one subgame at a time, we solve the subgame given the
current trunk strategy, update the trunk using the counterfactual
values at the root of the subgame, update the average counterfactual
values at the root of the subgame, and then discard the solution to
the subgame.  We then update the trunk using the current trunk
strategy.  The average strategy is an approximation of a Nash
equilibrium, where we don't know any action probabilities in the
subgames.  Note that we must keep the average counterfactual values at
the root of the subgames if we wish to use subgame re-solving to find
a policy in the subgame after solving.

\begin{theorem}
Let $\mathcal{I}_{TR}$ be the information sets in the trunk, $A = \max_{I \in \mathcal{I}} |A(I)|$ be
an upper bound on the number of actions, and $\Delta = \max_{s,t\in
Z}|u(s)-u(t)|$ be the variance in leaf utility.  Let $\sigma^t$ be the
current CFR-D strategy profile at time $t$, and $N_{S}$ be the number
of information sets at the root of any subgame.  If for all times $t$,
players $p$, and information sets $I$ at the root of a subgame
$\mathcal{I}_{SG}$, the quantity
$R^T_{full}(I) = \displaystyle \max_{\sigma'}
v^{\sigma^t_{[\mathcal{I}_{SG} \gets \sigma']}}_p(I) -
v^{\sigma^t}_p(I)$ is bounded by $\epsilon_S$, then player $p$
regret $R^T_p \le \Delta |\mathcal{I}_{TR}|\sqrt{AT} +
TN_{S}\epsilon_{S}$.
\end{theorem}

\begin{proof}
The proof follows from Zinkevich \etal's argument in Appendix
A.1~\cite{07nips-cfr}.  Lemma 5 shows that for any player $p$
information set $I$, $R^T_{full}(I) \le R^T(I) + \sum_{I' \in
Child_p(I)} R^T_{full}(I')$ where $Child_p(I)$ is the set of all
player information sets which can be reached from $I$ without passing
through another player $p$ information set.

We now use an argument by induction.  For any trunk information set
$I$ with no descendants in $\mathcal{I}_{TR}$, we have
$R^T_{full}(I) \le R^T(I) \le R^T(I) + TN_S\epsilon_S$.

Assume that for any player $p$ information set $I$ with no more than
$i \ge 0$ descendants in $\mathcal{I}_{TR}$,
$R^T_{full}(I) \le \sum_{I' \in Trunk(I)} R^T(I') + TN_{S}\epsilon_S$,
where $Trunk(I)$ is the set of player $p$ information sets in
$\mathcal{I}_{TR}$ reachable from $I$, including $I$.  Now consider a
player $p$ information set with $i+1$ descendants.  By Lemma 5 of
Zinkevich \etal, we get $R^T_{full} \le R^T(I) + \sum_{I' \in
Child_p(I)} R^T_{full}(I')$.  Because $I'$ must have no more than $i$
descendants for all $I' \in Child(I)$, we get
$R^T_{full}(I) \le \sum_{I' \in Trunk(I)} R^T(I') + TN_{S}\epsilon_S$.

By induction this holds for all $i$, and must hold at the root of the
game, so $R^T_p \le \sum_{I \in \mathcal{I}_{TR}} R^T(I') +
TN_{S}\epsilon_S$.  We do regret matching in the trunk, so
$R^T(I') \le \Delta \sqrt{AT}$ for all $I'$.
\end{proof}

The benefit of CFR-D is the reduced memory requirements.  CFR-D only
stores values for information sets in the trunk and at the root of
each subgame, giving it memory requirements which are sub-linear in
the number of information sets.  Treating the subgames independently
can lead to a substantial reduction in space: $\mathcal{O}(\sqrt{S})$
instead of $\mathcal{O}(S)$, as described in the introduction.  There
are two costs to the reduced space.  The first is that the subgame
strategies must be re-solved at run-time.  The second cost is
increased CPU time to solve the game.  At each iteration, CFR-D must
find a Nash equilibrium for a number of subgames.  CFR variants
require $\mathcal{O}(1/\epsilon^2)$ iterations to have an error less
than $\epsilon$, and this bound applies to the number of trunk
iterations required for CFR-D.  If we use CFR to solve the subgames,
each of the subgames will also require $\mathcal{O}(1/\epsilon^2)$
iterations at each trunk iteration, so CFR-D ends up doing
$\mathcal{O}(1/\epsilon^4)$ work.

In CFR-D, the subgame strategies must be mutual counterfactual best
responses, not just mutual best responses.  The only difference is
that a counterfactual best response will maximise counterfactual value
at an information set $I$ where $\pi_{P(I)}(I) = 0$.  A best response
may choose an arbitrary policy at $I$.  While CFR naturally produces a
mutual counterfactual best response, a subgame equilibrium generated by
some other method like a sequence form linear program may not be a
counterfactual best response.  In this case, the resulting strategy
profile is easily fixed with a post-processing step which computes the
best response using counterfactual values whenever $\pi^\sigma_p(I)$
is 0.

\section{Experimental Results}
We have three main claims to demonstrate.  First, if we have a
strategy, we can reduce space usage by keeping only summary
information about the subgames, and then re-solve any subgame with
arbitrarily small error.  Second, we can decompose a game, only use
space for the trunk and a single subgame, and generate an arbitrarily
good approximation of a Nash equilibrium using CFR-D.  Finally, we can
use the subgame re-solving technique to reduce the exploitability of
an existing strategy.  All results were generated on a 2.67GHz Intel
Xeon X5650 based machine running Linux.

\subsection{Re-Solving Strategies in Subgames}
To show that re-solving subgames introduces at most an
arbitrarily small exploitability, we use
the game of Leduc Hold'em poker, a popular research testbed for
imperfect information
games \cite{09aamas-abstraction,GanzfriedSW11}.  
The game uses a 6-card deck and has two betting rounds, with 936 information sets total.  It retains interesting strategic elements while being small enough that a range of experiments can be easily run and evaluated.
In this experiment, the trunk used was the first round of betting, and
there were five subgames corresponding to the five different betting
sequences where no player folds.  When re-solving for
subgame strategies, we used the Public Chance Sampling (PCS)
 variant of CFR~\cite{12aamas-pcs}.

To demonstrate the practicality of re-solving subgame strategies, we
started with an almost exact Nash equilibrium (exploitable by less
than $2.5*10^{-11}$ chips per hand), computed the counterfactual
values of every hand in each subgame for both players, and discarded
the strategy in all subgames.  These steps correspond to a
real scenario where we pre-compute and store a Nash equilibrium in
an offline fashion.  At run-time, we then re-solved each subgame
using the subgame re-solving game constructed from the
counterfactual values and trunk strategy, and measured the
exploitability of the combined trunk and re-solved subgame strategies.

\begin{figure}
\begin{tabular}{c}
\includegraphics[width=0.8\hsize]{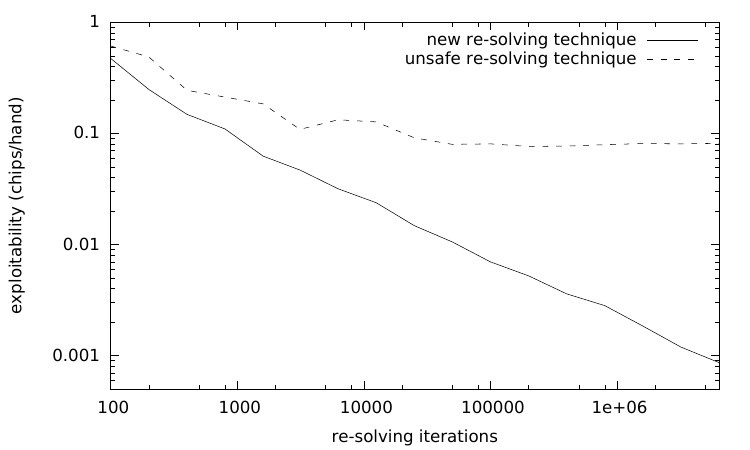} \\
\end{tabular}
\caption{exploitability after subgame re-solving}
\label{fig:leduc-recovery}
\end{figure}

Figure~\ref{fig:leduc-recovery} shows the exploitability when using a
different number of CFR iterations to solve the re-solving games.  The
$\mathcal{O}(1/\sqrt{T})$ error bound for CFR in the re-solving games
very clearly translates into the expected $\mathcal{O}(1/\sqrt{T})$
error in the overall exploitability of the re-constructed strategy.

For comparison, the ``unsafe re-solving technique'' line in
Figure~\ref{fig:leduc-recovery} shows the performance of a system for
approximating undominated subgame
solutions~\cite{GanzfriedSandholm13}.  Not only is there no
theoretical bound on exploitability, the real world behaviour is not
ideal.  Instead of approaching a Nash equilibrium (0 exploitability),
the exploitability of the re-solved strategy approaches a value of
around 0.080 chips/hand.  Re-solving time ranged from 1ms for 100
iterations, up to 25s for 6.4 million iterations, and the safe
re-solving method was around one tenth of a percent slower than unsafe
re-solving.

\subsection{Solving Games with Decomposition}
\begin{figure}
\begin{tabular}{c}
\centering\includegraphics[width=0.8\hsize]{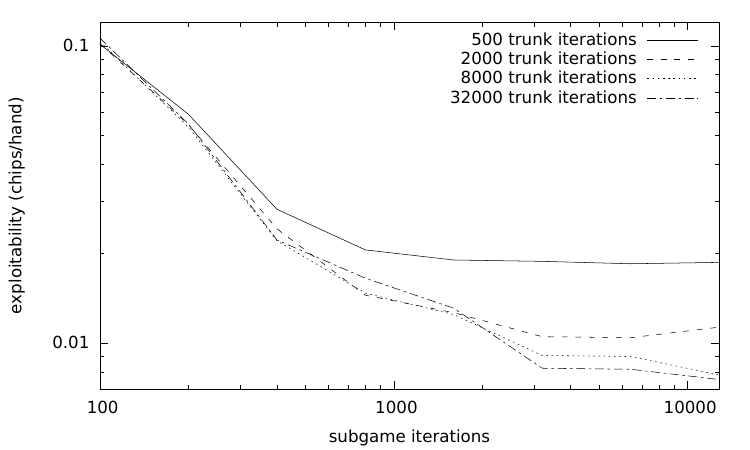}
\end{tabular}
\caption{CFR-D exploitability}
\label{fig:leduc-results}
\end{figure}

To demonstrate CFR-D, we split Leduc Hold'em in the same fashion as
the strategy re-solving experiments.  Our implementation of CFR-D used
CFR for both solving subgames while learning the trunk strategy and the 
subgame re-solving games.  All
the reported results use 200,000 iterations for each of the re-solving
subgames (0.8 seconds per subgame.)  Each line of
Figure~\ref{fig:leduc-results} plots the exploitability for different
numbers of subgame iterations performed during CFR-D, ranging from 100 to 12,800 iterations.
There are results for 500, 2,000, 8,000, and 32,000 trunk iterations.

Looking from left to right, each of the lines show the decrease
in exploitability as the quality of subgame solutions increases.  The
different lines compare exploitability across an increasing number of
CFR-D iterations in the trunk.

Given that the error bound for CFR variants is
$\mathcal{O}(\sqrt{T})$, one might expect exploitability results to be
a straight line on a log-log plot.  In these experiments,
CFR-D is using CFR for the trunk, subgames, and the re-solving games, so
the exploitability is a sum of trunk, subgame, and subgame
re-solving errors. For each line on the graph, trunk
and subgame re-solving error are constant values.  Only subgame error
decreases as the number of subgame iteration increases, so each line
is approaching the non-zero trunk and re-solving error, which shows up
as a plateau on a log-log plot.

\subsection{Re-Solving to Improve Subgame Strategies}
Generating a new subgame strategy at run-time can also be used to
improve the exploitability of a strategy.  In large games, lossy abstraction
techniques are often used to reduce a game to a tractable
size~\cite{13aamas-abstraction}.  When describing their recent subgame solving
technique, Ganzfried~\etal~reported positive results in experiments where
subgames are re-solved using a much finer-grained abstract game than the
original solution~\cite{GanzfriedSandholm13}.  Our new subgame re-solving
method adds a theoretical guarantee to the ongoing research in this
area.

\begin{figure}
\begin{tabular}{c}
\includegraphics[width=0.8\hsize]{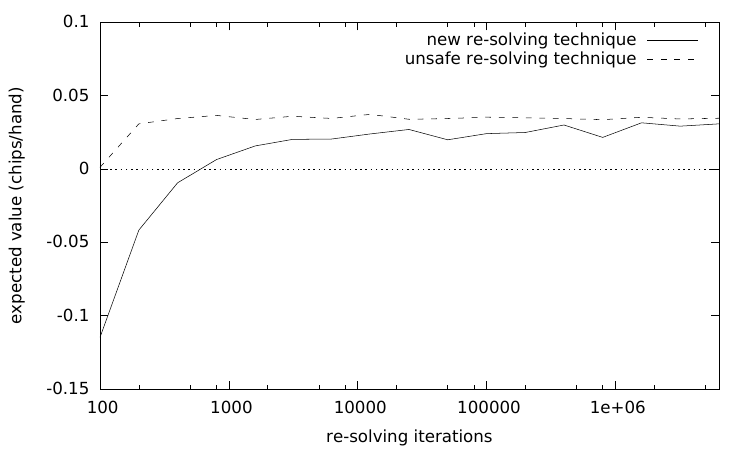} \\
\includegraphics[width=0.8\hsize]{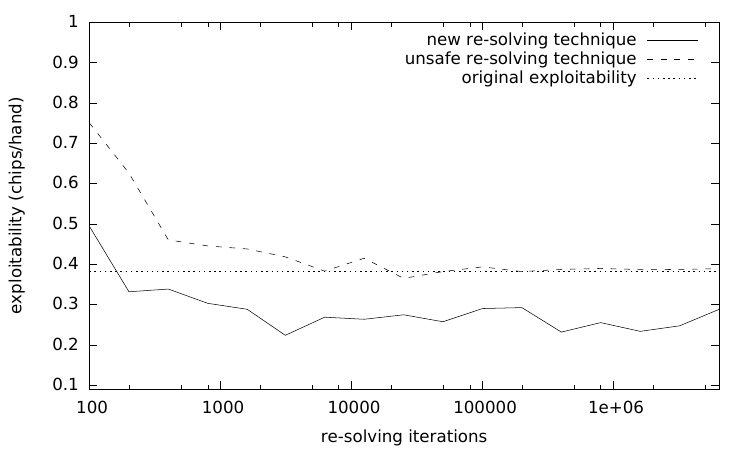}
\end{tabular}
\caption{plots of re-solved abstract strategy in Leduc Hold'em showing performance against the original strategy (top) and exploitability (bottom)}
\label{fig:bucketed-leduc-resolve}
\end{figure}

In Figure~\ref{fig:bucketed-leduc-resolve}, we demonstrate re-solving
subgames with a Leduc Hold'em strategy generated using an abstraction.  In
the original strategy, the players can not tell the difference between
a Queen or a King on the board if they hold a Jack, or between a 
Jack or a Queen on the board if they hold a King.  This abstraction 
gives a player perfect knowledge of the strength of their hand against
a uniform random hand, but loses strategically important ``textural'' information
and the resulting strategy is exploitable for 0.382 chips/hand in the full game.  
To generate the
counterfactual values needed for our method, we simply do a best
response computation within the subgame: the standard recursive best
response algorithm naturally produces counterfactual values.

The top plot shows the expected value of an abstract trunk strategy with
re-solved unabstracted subgames, when played in the full game 
against the original abstract strategy.  
With little effort, both re-solving techniques see
some improvement against the original strategy.  With more effort, the
unsafe re-solving technique has a small edge of around 0.004
chips/hand over our re-solving technique.

The bottom plot measures the exploitability of the re-solved
strategies.  Within 200 iterations, our new re-solving method
decreases the 
exploitability to 0.33 chips/hand.  After 2,000
iterations, the exploitability ranges between 0.23 and 0.29
chips/hand.  The unsafe method, after 6,250 iterations, stays at 0.39
chips/hand.  Note that the unsafe method's apparent convergence to the
original exploitability is a coincidence: in other situations the
unsafe strategy can be less exploitable, or significantly
more exploitable.

If we have reliable information about the opponent's trunk strategy, we might
want to use the unsafe re-solving method for its slight advantage in 
one-on-one performance.
Otherwise, the large difference in exploitability between the re-solving methods
supports our safe re-solving method.  This produces a robust strategy with 
a guarantee that the re-solved strategy does no worse than the original 
strategy.

\section{Conclusions}
In perfect information games, decomposing the problem into
independent subgames is a simple and effective method which
is used to greatly reduce the space and time requirements of
algorithms.  It has previously not been known how to decompose imperfect
information domains without a loss of theoretical guarantees on
solution quality.  We present a method of using summary information
about a subgame strategy to generate a new strategy which is no more
exploitable than the original strategy.  Previous methods have no
guarantees, and we demonstrate that they produce strategies which can
be significantly exploitable in practice.

We also present CFR-D, an algorithm which uses decomposition to solve
games.  For the first time, we can use decomposition to achieve
sub-linear space costs, at a cost of increased computation time.
Using CFR-D, we can solve 2-Player Limit Texas Hold'em Poker in less
than 16GB, even though storing a complete strategy would take over
200TB of space.  While the time cost of solving Limit Hold'em is
currently too large, this work overcomes one of the key barriers to
such a computation being feasible.

\section{Acknowledgements}
This research was supported by the Natural Sciences and Engineering
Research Council (NSERC), Alberta Innovates Centre for Machine
Learning (AICML), and Alberta Innovates Technology Futures (AITF).
Computing resources were provided by Compute Canada.

\bibliography{cfrd}
\bibliographystyle{aaai}

\ifarxiv
\appendix
\section{Appendix: Proofs}
\label{sec:proofs}
First, we show that if we re-solve a subgame for player one, and do
not increase player two's best response counterfactual values, the
exploitability of the combined player one trunk and subgame strategy
is no higher than the exploitability of the original player one
strategy.

{
\renewcommand{\thetheorem}{\ref{theorem:resolve-subgames}}
\begin{theorem}
Given a strategy $\sigma_1$, a subgame $S$, and a
re-solved subgame strategy $\sigma^S_1$, let $\sigma'_p
= \sigma_{p,[S \gets \sigma^S_p]}$ be the combination of $\sigma_1$
and $\sigma^S_1$. If
$v_2^{\langle\sigma'_1,\CBR(\sigma'_1)\rangle}(I) \le
v_2^{\langle\sigma_1,\CBR(\sigma_1)\rangle}(I)$ for all information
sets $I$ at the root of subgame $S$, then
$u_2^{\langle\sigma'_1,\CBR(\sigma'_1)\rangle}
\le u_2^{\langle\sigma_1,\CBR(\sigma_1)\rangle}$.
\end{theorem}
\addtocounter{theorem}{-1}
}

\begin{proof}
By definition, the strategy $\CBR(\sigma'_1)$ maximises counterfactual
value $v^{\langle \sigma',CBR(\sigma'_1) \rangle}_2(I)$ at all player
two information sets.  Because $v_2(I)$ does not depend on the player
two strategy before $I$, we can construct $\CBR(\sigma'_1)$
recursively, so that $CBR(\sigma'_1)(I)
= \argmax_{a \in A(I)}v^{\langle \sigma'_1, CBR(\sigma'_1) \rangle}_2(I\cdot
a)$.

By assumption, we have
$v_2^{\langle\sigma'_1,\CBR(\sigma'_1)\rangle}(I) \le
v_2^{\langle\sigma_1,\CBR(\sigma_1)\rangle}(I)$ for any information
set $I$ at the root of the subgame $S$.  Because
$\sigma'_1(I',a)=\sigma_1(I',a)$ for any information set $I'$ outside
$S$, for any information set $I$ which can not reach $S$,
$v_2^{<\sigma'_1,\CBR(\sigma'_1)>}(I)$ does not depend on the policy
at any information set where $\sigma'_1$ differs from $\sigma_1$,
and $v_2^{\langle\sigma'_1,\CBR(\sigma'_1)\rangle}(I) =
v_2^{\langle\sigma_1,\CBR(\sigma_1)\rangle}(I)$.

Let us say $\depth(I) = 0$ if $I$ is at the root of subgame $S$ or $I$
can not reach $S$, and $\depth(I) = \max_{I' \in
child(I)}\depth(I')+1$ otherwise.  Assume that
$v_2^{\langle\sigma'_1,\CBR(\sigma'_1)\rangle}(I) \le
v_2^{\langle\sigma_1,\CBR(\sigma_1)\rangle}(I)$ for any information
set $I$ with $\depth(I) \le i$ for some $i \ge 0$.  From above, we
know this is true for $i=0$.

Now consider an information set $I$ with depth $i+1$.  By the
recursive definition of $\CBR()(I)$,
$v_2^{\langle\sigma'_1,\CBR(\sigma'_1)\rangle}(I) = max_{a \in
A(I)}v^{\langle \sigma'_1, CBR(\sigma'_1) \rangle}_2(I\cdot a)$ and
$v_2^{\langle\sigma_1,\CBR(\sigma_1)\rangle}(I) = max_{a \in
A(I)}v^{\langle \sigma_1, CBR(\sigma_1) \rangle}_2(I\cdot a)$.

$depth(I \cdot a) < depth(I)$, so
$depth(I \cdot a) \le i$, and by assumption $v^{\langle \sigma'_1,
CBR(\sigma'_1) \rangle}_2(I\cdot a)
\le v^{\langle \sigma_1, CBR(\sigma_1) \rangle}_2(I\cdot a)$ for all $a$.
The inequality must then hold for the maximum, and we have
$v_2^{\langle\sigma'_1,\CBR(\sigma'_1)\rangle}(I) \le
v_2^{\langle\sigma_1,\CBR(\sigma_1)\rangle}(I)$.

By induction, this must hold for all $i$, and so
$v_2^{\langle\sigma'_1,\CBR(\sigma'_1)\rangle}(I) \le
v_2^{\langle\sigma_1,\CBR(\sigma_1)\rangle}(I)$ for an information set
$I$ at the root of the game.  If no player two actions have yet been
taken in the game at $I$, $v_2(I)=u_2(I)$, and
$u_2^{\langle\sigma'_1,\CBR(\sigma'_1)\rangle}
\le u_2^{\langle\sigma_1,\CBR(\sigma_1)\rangle}$.
\end{proof}

Next, Theorem~\ref{theorem:resolved-cfrd} gives a proof of the upper bound
on exploitability of a recovered strategy.  The context for this
section is as follows.  Strategy profile $\sigma$ is an approximation
of a Nash equilibrium for the whole game.  The induced recovery game
strategy profile $\sigma^{\tilde{F}}$ is the strategy where for all
information sets in the subtrees under the $F$ action,
$\sigma^{\tilde{F}}$ takes the same action as $\sigma$, and at the
$p_2$ information sets where $F$ or $T$ is chosen, $p_2$ always picks
$F$.  We will be considering the process from the point of view of
recovering a strategy for $p_1$.

\begin{lemma}
\label{lemma:switching_games}
For any $p_2$ strategy $\rho$ in the original game and $p_1$ strategy
$\tilde{\rho}$ in the recovery game, if we let $\hat{\sigma} = \langle
\sigma_{1[SG \gets \tilde{\rho}]}, \rho \rangle$, then for any $I \in
\mathcal{I}^{\tilde{R}}_2$, $u_2^{\hat{\sigma}}(I) = \pi_{2}^{\rho}(I)
\tilde{u}_2^{\langle \tilde{\rho}, \rho^{\tilde{F}} \rangle}(I)$.

\begin{proof}
\begin{align}
  u_2^{\hat{\sigma}}(I) \nonumber \\
 = \sum_{z \in Z(I)} \pi_{2}^{\rho}(z[I]) \pi_{-2}^{\sigma}(z[I]) \pi_{2}^{\rho}(z[I],z) \pi_{-2}^{\tilde{\rho}}(z[I],z) u_2(z) \nonumber \\
= \pi_{2}^{\rho}(I) \sum_z \pi_{-2}^{\sigma}(z[I]) / k * \pi_{2}^{\rho}(z[I],z) \pi_{-2}^{\tilde{\rho}}(z[I],z) u_2(z)k \nonumber \\
= \pi_{2}^{\rho}(I) \tilde{u}_2^{\langle \tilde{\rho}, \rho^{\tilde{F}} \rangle}(I) \nonumber
\end{align}
\end{proof}
\end{lemma}

\begin{lemma}
\label{lemma:inequality}
If $\tilde{\sigma}$ is an $\epsilon_R$-Nash equilibrium in the
recovery game, $0 \le c_I \le 1$, and $u_2^{\langle \sigma_1,
  BR(\sigma_1) \rangle}(I) \le \epsilon_S + u_2^{\sigma}(I)$ for all
$I$, then
\begin{align}
  \sum_{I \in \mathcal{I}_2^{\tilde{R}}}c_I\tilde{u}_2^{\langle \tilde{\sigma}_1, BR(\tilde{\sigma}_1) \rangle}(I) \nonumber \\
\le (|\mathcal{I}_2^{\tilde{R}}|-1)\epsilon_S + \epsilon_R + \sum_I{c_I\tilde{u}_2^{\langle \sigma^{\tilde{F}}_1, BR(\sigma^{\tilde{F}}_1) \rangle}}(I) \nonumber
\end{align}

\begin{proof}
$\sigma$ and $\tilde{\sigma}$ have the following properties.
\begin{align}
\tilde{u}_2^{\langle \tilde{\sigma}_1, BR(\tilde{\sigma}_1) \rangle}
\le \epsilon_R + \tilde{u}_2^{\tilde{\sigma}^{*}}
\le \epsilon_R + \tilde{u}_2^{\langle \sigma^{\tilde{F}}_1, BR(\sigma^{\tilde{F}}_1) \rangle} \nonumber \\
v^\sigma_2(I)
\le \tilde{u}_2^{\langle \tilde{\sigma}_1, BR(\tilde{\sigma}_1) \rangle}(I)\nonumber \\
v^\sigma_2(I)
\le \tilde{u}_2^{\langle \sigma^{\tilde{F}}_1, BR(\sigma^{\tilde{F}}_1) \rangle}(I)
\le \epsilon_S + \tilde{u}_2^{\sigma^{\tilde{F}}}(I)
= \epsilon_S + v_2^{\sigma}(I) \nonumber
\end{align}
Given this, the maximum difference between $c \cdot
\tilde{u}_2^{\langle \tilde{\sigma}_1, BR(\tilde{\sigma}_1) \rangle}$ and
$c \cdot \tilde{u}_2^{\langle \sigma^{\tilde{F}_1},
  BR(\sigma^{\tilde{F}}_1) \rangle}$ occurs when the difference of these
sums is concentrated at a single $I$.  That is, for some $I$
\begin{align}
\tilde{u}_2^{\langle \tilde{\sigma}_1, BR(\tilde{\sigma}_1) \rangle}(I) = (|\mathcal{I}_2^{\tilde{R}}|-1)\epsilon_S +
\tilde{u}_2^{\langle \sigma^{\tilde{F}}_1, BR(\sigma^{\tilde{F}}_1) \rangle}(I) \nonumber \\
c_I=1 \nonumber
\end{align}
and for all $I' \ne I$
\begin{align}
\tilde{u}_2^{\langle \tilde{\sigma}_1, BR(\tilde{\sigma}_1) \rangle}(I')
= \tilde{u}_2^{\sigma^{\tilde{F}}}(I') \nonumber \\
\tilde{u}_2^{\langle \sigma^{\tilde{F}}_1, BR(\sigma^{\tilde{F}}_1) \rangle}(I')
= \epsilon_S + \tilde{u}_2^{\sigma^{\tilde{F}}}(I') \nonumber \\
c_{I'} = 0 \nonumber
\end{align}
In this case, the difference is $(|\mathcal{I}_2^{\tilde{R}}|-1)\epsilon_S + \epsilon_R$.
\end{proof}
\end{lemma}

\begin{theorem}
\label{theorem:resolved-cfrd}
Let $\sigma$ be a equilibrium profile approximation, where
$\epsilon_S$ is an upper bound on the $p_2$ counterfactual regret so
that $R_2(I) \le \epsilon_S$ over all $I$ in
$\mathcal{I}_2^{\tilde{R}}$.  Let $\tilde{\sigma}$ be the recovered
strategy, with a bound $\epsilon_R$ on the exploitability in the
recovery game.  Then the exploitability of $\sigma$ is increased by no
more than $(|\mathcal{I}_2^{\tilde{R}}|-1)\epsilon_S + \epsilon_R$ if we use $\tilde{\sigma}$
in the subgame:
\begin{align}
u_2^{\langle \sigma_{1[SG \gets \tilde{\sigma}]},BR(\sigma_{1[SG \gets \tilde{\sigma}]}) \rangle} \nonumber \\
\le (|\mathcal{I}_2^{\tilde{R}}|-1) \epsilon_S + \epsilon_R + u_2^{\langle \sigma_1, BR(\sigma_1) \rangle} \nonumber
\end{align}

\begin{proof}
  Let $\hat{\sigma} = \langle \sigma_{1 [SG \gets
      \tilde{\sigma}_1]}, BR(\sigma_{1 [SG \gets
      \tilde{\sigma}_1]}) \rangle$.  In this case,
\begin{align}
u_2^{\hat{\sigma}} \nonumber \\
= \sum_{z \notin SG} \pi^{\hat{\sigma}}(z) u_2(z) + \sum_{z \in SG} \pi^{\hat{\sigma}}(z) u_2(z) \nonumber \\
= \sum_{z \notin SG} \pi^{\langle \sigma_1, BR(\sigma_1) \rangle}(z) u_2(z) + \sum_{z \in SG} \pi^{\hat{\sigma}}(z) u_2(z) \label{thm-line:orig-eqn}
\end{align}
Considering only the second sum, rearranging the terms and using
Lemma~\ref{lemma:switching_games}
\begin{align}
\sum_{z \in SG} \pi^{\hat{\sigma}}(z) u_2(z)
= \sum_{I \in \mathcal{I}_2^{R}} u^{\hat{\sigma}}_2(I)
= \sum_{I \in \mathcal{I}_2^{\tilde{R}}} \pi_{2}^{\hat{\sigma}}(I) \tilde{u}_2^{\langle \tilde{\sigma_1}, \hat{\sigma_2}^{\tilde{F}} \rangle}(I) \nonumber
\end{align}
A best response must have no less utility than $\hat{\sigma}^{\tilde{F}}_2$, and we can then apply Lemma~\ref{lemma:inequality}
\begin{align}
\le \sum_I \pi_{2}^{\hat{\sigma}}(I) \tilde{u}_2^{\langle \tilde{\sigma}_1, BR(\tilde{\sigma}_1) \rangle}(I) \nonumber \\
\le (|\mathcal{I}_2^{\tilde{R}}|-1) \epsilon_S + \epsilon_R + \sum_I \pi_{2}^{\hat{\sigma}}(I) \tilde{u}_2^{\langle \sigma^{\tilde{F}}_1, BR(\sigma^{\tilde{F}}_1) \rangle}(I) \nonumber
\end{align}
Because $\tilde{u}^{\sigma^{\tilde{F}}}_2(I) = v^\sigma_2(I)$ and
$\tilde{u}_2(I \cdot T) = v^\sigma_2(I)$ for all $I$,
$BR(\sigma^{\tilde{F}}_1)$ can always pick action $F$, and we can
directly use $BR(\sigma^{\tilde{F}}_1)$ in the real game, with the
same counterfactual value.
\begin{align}
= (|\mathcal{I}_2^{R}|-1) \epsilon_S + \epsilon_R + \sum_I \pi_{2}^{\hat{\sigma}}(I) v_2^{\langle \sigma_1, BR(\sigma_1) \rangle}(I) \nonumber
\end{align}
Putting this back into line~\ref{thm-line:orig-eqn}, and noting that a
best response can only increase the utility, we get
\begin{align}
u_2^{\hat{\sigma}}
= (|\mathcal{I}_2^{R}|-1) \epsilon_S + \epsilon_R + u^{\langle \sigma_1, \hat{\sigma}_{[SG \gets BR(\sigma)}\rangle}_2 \nonumber \\
\le (|\mathcal{I}_2^{\tilde{R}}|-1) \epsilon_S + \epsilon_R + u_2^{\langle \sigma_1, BR(\sigma_1) \rangle} \nonumber
\end{align}
\end{proof}
\end{theorem}
\fi
\end{document}